\def\gsim{~\rlap{$>$}{\lower 1.0ex\hbox{$\sim$}}}
\def\lsim{~\rlap{$<$}{\lower 1.0ex\hbox{$\sim$}}}

\def\h2o{\rm{H_{2}O}}
\def\mh2{\rm{H_{2}}}

\def\co2{\rm{CO_{2}}}
\def\ch4{\rm{CH_{4}}}
\def\nh3{\rm{NH_{3}}}
\def\n2{\rm{N_{2}}}

\documentclass[12pt, preprint]{aastex}
\usepackage{graphicx}
\usepackage{natbib,color,lscape}
\usepackage{subfigure}
\usepackage{threeparttable}
\usepackage{longtable}
\usepackage{url}
\usepackage{subfigure}
\usepackage{rotating}

\begin{document}
\title{Exoplanet Classification and Yield Estimates for Direct Imaging Missions}
\author{Ravi kumar Kopparapu\altaffilmark{1,2,3,4}, Eric H\'ebrard\altaffilmark{1,5}, Rus Belikov\altaffilmark{6},  Natalie M. Batalha\altaffilmark{6}, Gijs D. Mulders\altaffilmark{7,8}, Chris Stark\altaffilmark{9}, Dillon Teal\altaffilmark{1}, Shawn Domagal-Goldman\altaffilmark{1,3}, Avi Mandell\altaffilmark{1}}

\altaffiltext{1}{NASA Goddard Space Flight Center, 8800 Greenbelt Road, Mail Stop 699.0, Building 34, Greenbelt, MD 20771}
\altaffiltext{2}{Department of Astronomy, University of Maryland College Park, College Park, MD}
\altaffiltext{3}{NASA Astrobiology Institute's Virtual Planetary Laboratory, P.O. Box 351580, Seattle, WA 98195, USA}
\altaffiltext{4}{Blue Marble Space Institute of Science, 1001 4th Ave, Suite 3201, Seattle, Washington 98154, USA}
\altaffiltext{5}{School of Physics and Astronomy, University of Exeter, EX4 4QL, Exeter, UK}
\altaffiltext{6}{NASA Ames Research Center}
\altaffiltext{7}{Lunar and Planetary Laboratory, University of Arizona, Tucson, AZ 85721, USA}
\altaffiltext{8}{Earths in Other Solar Systems Team, NASA Nexus for Exoplanet System Science, USA}
\altaffiltext{9}{Space Telescope Science Institute, 3700 San Martin Drive, Baltimore, MD 21218 }

\begin{abstract}
Future NASA concept missions that are currently under study, like Habitable Exoplanet Imaging Mission (HabEx) \& 
Large Ultra-Violet Optical Infra Red (LUVOIR) Surveyor, would discover a large diversity of exoplanets.
We propose here a classification scheme that distinguishes exoplanets into different categories based on their size and incident 
stellar flux, for the purpose of providing the expected number of exoplanets observed (yield) with direct imaging missions. 
The 
boundaries of this classification can be computed using the known chemical behavior of gases and condensates at different pressures
and temperatures in a planetary atmosphere. In this study, we initially focus on condensation curves for sphalerite ZnS, $\h2o$, 
$\co2$ and $\ch4$. The order in which these species condense in a planetary atmosphere define the boundaries between different
classes of planets.  Broadly, the planets are divided into rocky ($0.5 - 1.0$R$_{\oplus}$), super-Earths 
($1.0 - 1.75$R$_{\oplus}$), sub-Neptunes ($1.75 - 3.5$R$_{\oplus}$), sub-Jovians ($3.5 - 6.0$R$_{\oplus}$) and 
Jovians ($6 - 14.3$R$_{\oplus}$) based on their planet sizes, and 'hot', 'warm' and 'cold' based on the incident stellar flux. We
then calculate planet occurrence rates within these boundaries for different kinds of exoplanets, $\eta_{planet}$, 
using the community co-ordinated results of NASA's Exoplanet 
Program Analysis Group's Science Analysis Group-13 (SAG-13). These occurrence rate estimates are in turn used to estimate the 
expected exoplanet yields for direct imaging missions of different telescope diameter. 
\end{abstract}

\maketitle

\section{Introduction}
The discoveries of exoplanets over the last two decades has revealed planetary bodies of various sizes and masses around other stars \citep{Rowe2014, Anglada-Escude2016, Coughlin2016,  Kane2016, Morton2016, Gillon2017, Dittman2017}. More specifically, the location of these exoplanets around their host star has a significant influence not only on the prospects of their detectability, but also on the atmospheric chemical composition and the capability of characterizing these atmospheres. Furthermore, the relatively large sample of nearly 3400 confirmed planets and another $\sim 4700$ planet candidates to-date\footnote{http://exoplanetarchive.ipac.caltech.edu/} has enabled us to calculate exoplanet occurrence rates in our galaxy \citep{CS2011, Traub2012, Howard2012, Bonfils2013, DC2013, Petigura2013, KoppOccr2013, Gaidos2013, Fressin2013, DZ2013, FM2014, MS2014, Silburt2015, DC2015, Burke2015, Mulders2015}. These initial estimates are dominated by close-in planets due to the sensitivity of the detection techniques and search pipelines. Nevertheless, these studies made a crucial and a significant leap in understanding planet diversity, and paved a way for comparative planetology of exoplanets. 

Several of the above mentioned studies have also focused on obtaining an estimate of the fraction of stars that have at least one terrestrial mass/size planet in the habitable zone (HZ), or $\eta_{\oplus}$. 
Estimates of $\eta_{\oplus}$ for Sun-like stars have been calculated by the data collected from the $Kepler$ mission. Earlier estimates ranged from 0.02 \citep{FM2014} to 0.22 \citep{Petigura2013} for GK dwarfs, but more recent analyses \citep{Burke2015} imply that systematic errors dominate.  
For M-dwarfs, $\eta_{\oplus}$ is estimated to be $\sim 20\%$ 
on an average \citep{DC2015}. Apart from the general curiosity of finding how common are Earth-like planets in our galaxy, the focus on $\eta_{\oplus}$  has a more practical application: It can be used in the design of direct imaging missions, like the concept studies under consideration HabEx\footnote{http://www.jpl.nasa.gov/habex/}, the Habitable Exoplanet Explorer; and LUVOIR\footnote{https://asd.gsfc.nasa.gov/luvoir/}, the Large UV-Optical-InfraRed surveyor
 with the goal of detecting biosignatures, and also in calculating `exo-Earth candidate yield', the number of potentially habitable extrasolar planets (exoEarth candidates) that can be detected and spectroscopically characterized \citep[e.g.,][]{Stark2014, Stark2015}.

Crucial to these estimates are the location of the main-sequence HZ, which have been studied by both 1-D and 3-D climate models \citep{Kasting1993, Selsis2007b, Abe2011, PG2011, Kopp2013, Leconte2013a, Yang2013, zsom2013, Kopp2014, WT2014, Yang2014a, WT2015, Way2015, Godolt2015, Leconte2015, Kopp2016, HM2016, RK2017, Kopp2017}. With some exceptions for certain types of planets \citep{Kane2014}, there has not been an overarching way to classify planets beyond the HZ. The lack of a systematic way to classify exoplanets in general, combined with the allure of planets within it, has led to direct imaging mission yield analyses that focus on HZ planets to the exclusion of everything else \citep[e.g.,][]{Stark2014, Stark2015}. While some mission studies have attempted to classify the non-HZ planets into hot, warm and cold planets that a mission would discover\footnote{https://exoplanets.nasa.gov/exep/studies/probe-scale-stdt/}, the boundaries for the classification are arbitrarily fixed without giving consideration to chemical behavior of gases and condensates in a planetary atmosphere.  Classifying different size planets  based on the  transition/condensation of different species 
\citep{Burrows2004, Burrows2005}
provides a physical motivation in estimating exoplanet mission yields, separate from exo-Earth candidate yields. 

In the search for exo-Earth candidates, we will undoubtedly detect a multitude of brighter planets. According to \cite{Stark2014}, for an 8m size telescope, the number of exo-Earth candidates detected is $\sim 20$ (see Fig. 4 in Stark et al. 2014), although this is strongly dependent on the value of $\eta_{\oplus}$. 
At the same time, the number of stars observed to detect these exo-Earth candidates is $\sim 500$. If we assume that, on an average, every star has a planet of some 
size \citep{Cassan2012, Suzuki2016}, then there are  $\sim 500$ exoplanets of all sizes that can be observed. Not considering the $\sim 20$ exo-Earth candidates, the bulk of the exo-planets will fall into `non-Earth' classification, without any distinguishing features between them. This provides an additional motivation to devise a scheme based on planetary size and corresponding atmospheric characteristics of exoplanets.   

Some work has been done at the theoretical level 
to derive the radiative response of an irradiated atmosphere
\citep{RC2012, Parmentier2014, RC2014, Parmentier2015}, but these analytical tools have not been either used to derive 
any general boundaries nor tied to planet occurence estimates,
 and were not designed with that intent in mind. This highlights the need for a theory-based system to classify planets beyond the habitable zone for the purposes of understanding the diversity of worlds future missions could explore. Such a system should be based on the properties we can measure today, primarily size and orbital information, and the boundaries in the scheme should divide planets with major differences in the properties that would be observable with current and future missions.

Fortunately, much of the theory needed for such a classification scheme already exists. There has been significant progress in understanding how the size of a planet is a major control on composition, and therefore on future observables \citep{Rogers2010a, Rogers2010b}. This includes work on the relationship between size, density, and bulk composition \citep{Fortney2007, WM2014, Rogers2015, Wolfgang2016, ChenKipping2017}.  
While the exact values for the boundaries of mass-radius change based on the specific analysis or theoretical technique, 
there is growing evidence of structure in the occurrence rate distribution that suggests compositional aggregates:
 1) small, rocky worlds whose bulk composition and behavior is dominated by Fe, Mg and Si species; 2) planets with
 Fe/Mg/Si cores but significant gas envelopes consisting of H/He, $\ch4$, $\nh3$ ices; and 3) gas giants whose bulk composition and behavior is dictated almost exclusively by its volatiles.

Similarly, there have been significant advances in our understanding of how the orbital separation of non-HZ planets could affect the chemical composition of the atmosphere \citep{Cahoy2010}. A constant theme across these studies is the influence of clouds. As a planet moves further from its host star, its atmosphere will cool and lead to condensation of progressively less volatile chemicals in the atmosphere. This condensation would create a cold trap and an associated cloud deck. The result of this is a significant change in spectral properties: as the condensing species would be trapped at or below the cloud deck, the cloud deck itself would absorb and scatter light, causing preferential sampling of the layers at or above the cloud deck. Multiple "onion-like" cloud decks can form as sequentially less volatile species condense at higher altitudes for planets with greater star-planet separation distances and correspondingly lower levels of incoming stellar flux. This process can be observed in detail in the gas giants of our own Solar System \citep{EH1972}, and has been postulated to be a driver for the atmospheric structure and observable properties of exoplanet atmospheres \citep{BS1999, Sudarsky2003, Burrows2004, Burrows2005, Fortney2005, Marley2007, Morley2013, WS2015, Wakeford2017}. We also note that the habitable zone itself has been defined in a manner consistent with this, as the instellations (stellar incident flux) at which liquid water clouds form but carbon dioxide clouds do not \citep{Abe2011}.

This prior work enables the construction of an overarching scheme for identifying classes of planets. This scheme could apply to all worlds, regardless of whether they are rocky or gaseous (or something in between). And it would be both based on the current observable properties, and prediction of major transitions in future observables. In short, this represents a comprehensive means of predicting the diversity yields of future planet characterization missions. Below we discuss in more detail how we simulate the processes underlying this classification scheme. These simulations define the boundaries between different planet classes, for which we calculate occurrence rates based on prior exoplanet detection missions. The occurrence rates allow us to simulate exoplanet yields - not just for HZ planets but for a diversity of different kinds of worlds. Finally, we close with a discussion of the caveats of this approach, and the implications of this scheme for future missions.

 \section{A New Classification Scheme}
        \label{sec2}

A planet size - and the relationship between its size and mass - appears to be primarily driven by volatile inventory. For example, the atmospheric composition of larger planets is predominantly  H$_{2}$/He, while smaller planets can have a mixture of $\ch4$, $\co2$, $\h2o$ and $\nh3$. High-temperature atmospheres, such as hot Jupiters, should have their chemistry - and therefore their spectral features - determined primarily by equilibrium chemistry.  Low-temperature atmospheres will have chemistry dictated by photochemistry, but this will be secondary to determining what species are condensing in their atmospheres. The exception to this - which we will discuss later - is for photochemical aerosols, which could have a major impact in the same manner that clouds do.\\

The chemical behavior of gases and condensates in a planetary atmosphere can be determined as a function of pressure, temperature, and metallicity. Using results from Lodders and Fegley (2002) and Visscher et al. (2006) adapted with solar abundances taken from Lodders (2010), we have computed the condensation curves for sphalerite ZnS, $\h2o$, $\co2$ and $\ch4$ as a function of pressure and temperature for systems with a solar metallicity. 
Pressure-temperature profiles of planetary atmospheres are tightly related to the incoming stellar flux. We define the boundaries between our different selected planetary cases as the stellar fluxes for which these four species condense out.
For instance, ZnS clouds have been considered as possible condensates in hot exoplanet atmospheres \citep{Morley2012, Charnay2015}, so the location (or the stellar flux) at which ZnS clouds form in a planetary atmosphere denotes the first boundary for our planet classification. Moving further away from the star, at relatively lower stellar fluxes, $\h2o$ starts condensing in the atmosphere. This, then, becomes the next boundary for grouping different planets. Between these two boundaries is where one would expect to find ZnS mineral clouds and $\h2o$ in a gaseous state. Continuing to lower fluxes, $\co2$ and $\ch4$ condensates bracket final boundaries. The results are independent of any particular model atmospheres, and in principle, any pressure-temperature profile may be superimposed on these condensation curves to find the equilibrium composition along the profile. In particular, the intersection of a particular pressure-temperature profile with one of these condensation curves indicates the pressure and temperature at which the respective species condenses out in the planetary atmosphere considered. We have investigated the different incident stellar fluxes, or instellations, for which condensates could form within the regions of an exoplanet system that will be probed by future direct imaging missions. Specifically, we simulate the star-planet separations for which ZnS, $\h2o$, $\co2$ and $\ch4$ would condense out in planetary atmospheres. Other metallic clouds can condense at distances closer to the star than the ZnS condensation line, and other volatiles (e.g., NH$_{3}$) can condense at orbiting distances beyond the $\ch4$ condensation line, but such worlds are likely undetectable by future direct imaging missions so are not simulated here.

 We have considered six different planetary size boundaries: 0.5 R$_{\oplus}$, 1.0 R$_{\oplus}$, 1.75 R$_{\oplus}$, 3.5 R$_{\oplus}$, 6.0 R$_{\oplus}$, and 14.3 R$_{\oplus}$ in our grid. 
These boundaries represent, respectively: the radius (0.5 R$_{\oplus}$) at which planets in the habitable zone appear to not have a sufficient gravity well to retain atmospheres \citep{ZC2017};
the ``super-Earths'' ($1 - 1.75$ R$_{\oplus}$) and ``sub-Neptunes'' ($1.75 - 3.5$ R$_{\oplus}$), as defined by
\cite{Fulton2017} (see section 4.4) based on the observed gap in the radius distribution of small planets 
with orbital periods shorter than 100 days;
the assumed upper limit on Neptune-size planets (6R$_{\oplus}$) based on the small peak in the radius distribution
from \cite{Fulton2017};
and the radius past which planets transition to brown dwarf stars \citep{ChenKipping2017}. We have computed the 
corresponding pressure-temperature atmospheric profiles using (1) the non-grey analytical model of 
Parmentier and Guillot (2014) with the coefficients from Parmentier et al. (2015) and the Rosseland opacity functional 
form of Valencia et al. (2013), and (2) the grey analytical model of Robinson and Catling (2012, 2014), both modified 
to take the planetary size and instellation as unique input parameters. We have used the 
Robinson and Catling (2012, 2014) model for planets with radius smaller than 3.5 R$_{\oplus}$ at low instellations, 
and the Parmentier and Guillot (2014) model for planets with a radius of 14.3 R$_{\oplus}$ at high instellations. We 
have assumed an internal temperature $T_{int}=0$ and mass-radius relations taken from Weiss and Marcy (2014) for 
planets with radius smaller than 3.5 R$_{\oplus}$. We have assumed an internal temperature $T_{int}=100$ and the 
density calculated from Mass-radius relation from \cite{ChenKipping2017} for planets with radius equal to 
14.3 R$_{\oplus}$. Assuming a planetary mass of $0.414$M$_{J} (131$M$_{\oplus}$), the density is $\sim 0.25$
g.cm$^{-3}$.  We have 
considered $L = 0.95 L_{\odot}$ and $M = 0.965M_{\odot}$ for the parent star\footnote{The values of the stellar
luminosity and mass are obtained as follows: We downloaded the confirmed and candidate Kepler catalog from NEXSCI,
 found the median values of stellar luminosity and mass for each data set, and then took the average value of 
luminosity and mass from these median values.
Median Luminosity for candidate planet list: 1.057;
Median Stellar mass for candidate planet list: 0.97;
Median Luminosity for confirmed planet list: 0.86;
Median Stellar mass for confirmed planet list: 0.96;
Average luminosity of confirmed \& candidate: (1.057 + 0.86)/2 = 0.95;
Average stellar mass of confirmed \& candidate: (0.97+0.96)/2 = 0.965}.

\begin{figure}[!hbp]
\centering
\includegraphics[width=0.85\textwidth]{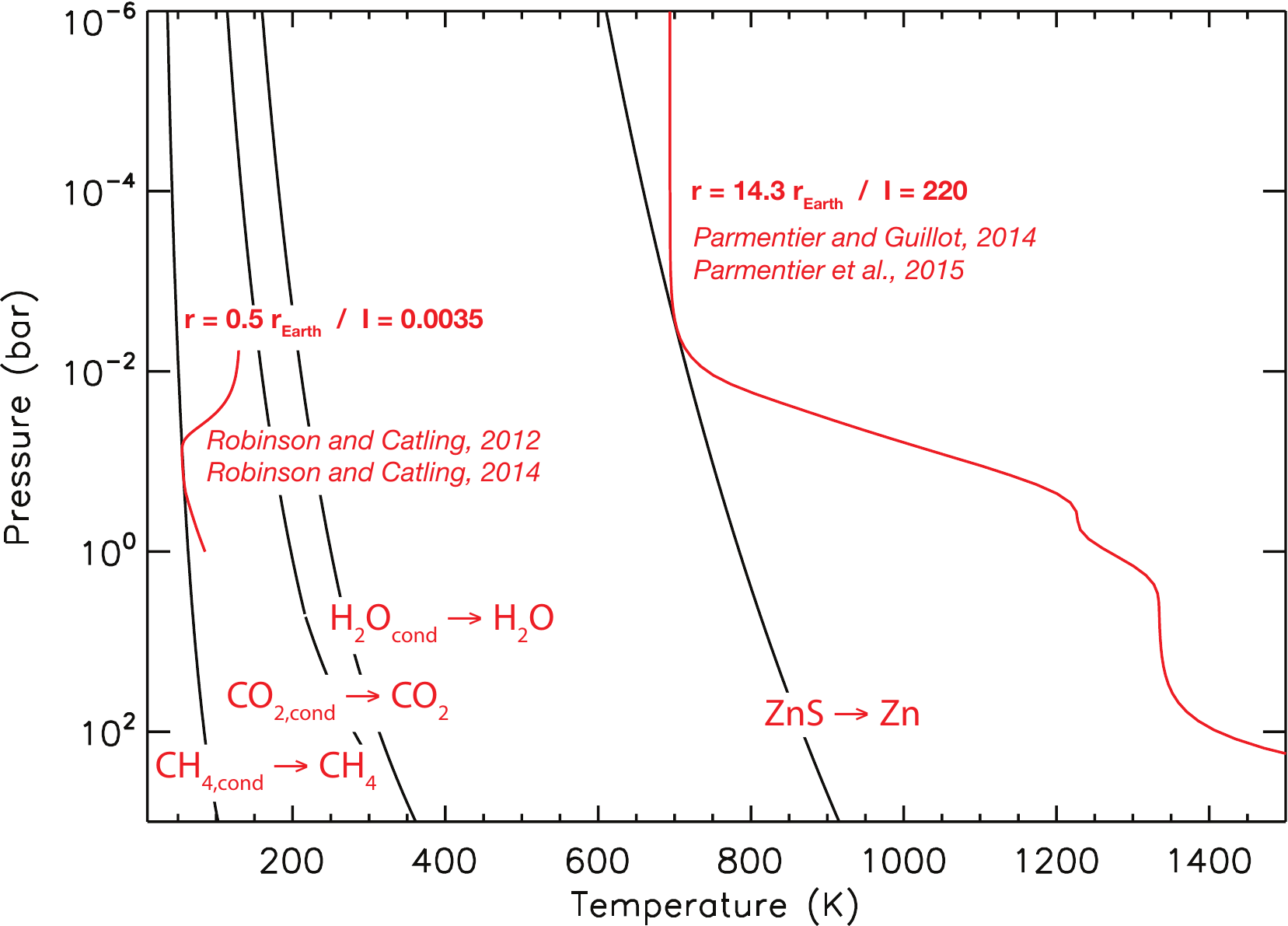}
\caption{Dependence of ZnS, $\h2o$, $\co2$ and $\ch4$ condensation with pressure and temperature in any planetary atmosphere (solid black) along with the pressure-temperature profiles for two different sizes of planets, 0.5 R$_{\oplus}$ and 14.3 R$_{\oplus}$ and two different instellations, 0.004 I$_{\oplus}$ and 220 I$_{\oplus}$, respectively (solid red). The intersections of the two sets of curves indicate that $\ch4$ and ZnS are condensing out in each of considered planetary atmospheres.}
\label{scheme}
\end{figure}

As an illustrative example, we show in Fig. \ref{scheme} the condensation curves for the four different species we 
considered (solid black), along with temperature profiles for two different size planets: $0.5$R$_{\oplus}$ and 
$14.3$R$_{\oplus}$, at different incident stellar fluxes (`instellation'). The figure shows that ZnS would condense 
out at $\sim10$mb in the atmosphere of a highly irradiated (220 times Earth flux) 14.3 Earth radius planet, a typical 
hot-Jupiter, whereas $\ch4$ would condense out in the atmosphere of a $0.5$  Earth radius planet receiving only 
$ \sim 1/280^{th}$ the flux Earth is receiving. Following this procedure, we have derived the radius and stellar flux 
where other gaseous species condense in the atmosphere. Table 1 provides the corresponding data of the planetary 
radius and stellar flux boundaries that can be used for classifying planets into different regimes.  These
boundaries are parameterized in Eq.(\ref{boundaries}) and are also available as an online calculator at: 
\url{http://www3.geosc.psu.edu/~ruk15/extrasolar/}.

 \begin{threeparttable}[h!]
\caption{ Planetary radii and stellar flux values at which the given species condense in a planetary atmosphere, 
for a star with $L = 0.95 L_{\odot}$ and $M = 0.965M_{\odot}$. These 
limits form the boundaries for classifying planets into different categories to calculate exoplanet yield estimates. See Fig. 2 and Section 2.2.}
\vspace{0.1 in}
\centering
\begin{tabular}{ |c|cccc| }
 \hline
  && Stellar Flux (Earth flux) &&  \\
  \hline
 Radius (R$_{\oplus}$) & ZnS  & $\h2o$ & $\co2$ & $\ch4$\\ 
 \hline
 0.5&182&1.0&0.28&0.0035\\
 1.0&187&1.12&0.30&0.0030\\
 1.75&188&1.15&0.32&0.0030\\
 3.5&220&1.65&0.45&0.0030\\
 6.0&220&1.65&0.40&0.0025\\
 14.3&220&1.7&0.45&0.0025\\
 \hline
\end{tabular}
\label{Table1}
\end{threeparttable}

\begin{eqnarray}
F(R_{p})_{i} &=& a_{i} x^{5} + b_{i} x^{4} + c_{i} x^{3} + d_{i} x^{2} + e_{i} x + f_{i}
\label{boundaries}
\end{eqnarray}

where $F(R_{p})_{i}$ is the stellar flux, normalized to the current Earth flux ($1360$ Wm$^{-2}$), at which species
$i =$ (ZnS, $\h2o, \co2, \ch4)$ condenses on a planet with radius $R_{p}$, and $x = R_{p}/R_{\oplus}$ given in
Table \ref{Table1}. The 
coefficients for the four condensing species are given in Table \ref{coeffs}.

\begin{threeparttable}[h!]
\caption{ Coefficients to be used in Eq.(\ref{boundaries}).}
\vspace{0.1 in}
\centering
\begin{tabular}{ |c|c|c|c|c|c|c| }
 \hline
  &a&b&c&d&e&f  \\
  \hline
 ZnS & 0.0010338041 &- 0.0255230451 & 0.1858822989 & - 0.4990171468 & 0.5690844110 & 1.6385396777 \\
 \hline
 $\h2o$ & 0.0017802416 & - 0.0443704003 & 0.3302966853 & - 0.9299682851 & 1.1366785108 & 0.6255832476 \\
 \hline
$\co2$ & 0.0002947546 & - 0.0070583509 & 0.0483928147 & - 0.1198359100 & 0.1477297602 & 0.2304769313 \\
 \hline
$\ch4$ & -0.0000033096 & 0.0000889715 & -0.0007644190 & 0.0026719183 & -0.0038305535 & 0.0048373922 \\
 \hline
\end{tabular}
\label{coeffs}
\end{threeparttable}

\subsection{Application of the Classification Scheme to Obtain Planet Occurrence Rates}
\label{Occurrence}

Extending the insights obtained from Fig. \ref{scheme} and Table 1, it is then possible to define various ‘zones’ as a function of stellar flux and planetary radius. This should qualitatively apply across all stellar types and the entire field of planets, even if the quantitative positions of the boundaries change due to, for example, the age of the system and the amount of internal heat released from planets, or the stellar energy distribution (SED) and its relationship to planetary albedo \citep{Segura2003}. For a full consideration of such caveats, see section 3.1. This particular framing of the parameter space will allow us to calculate the occurrence of different kinds of planets within each zone based on their condensation conditions, as both radius and flux are measurable quantities.

We have performed such a calculation of $\eta_{planet}$, the fraction of stars that have a planet within one of the zones defined by a condensing species. 
As an illustrative example of how these boundaries can be used to calculate the occurrence of planets, we apply these criteria to the preliminary parametric model introduced by one of NASA's Exoplanet Program Analysis Group (ExoPAG) science analysis studies (SAG13). A detailed discussion of the SAG13 model is outside the scope of this paper, but we will summarize the most critical points.

The SAG13 model is based on a simple meta-analysis of planet occurrence rates from many different individual publications and groups. Specifically, the SAG13 group collected tables of occurrence rates calculated over a standard grid of 
planet radius, period, and stellar type. A full description of the grid is as follows.
 The $i^{th}$ bin in the planet radius is defined as the interval:

\begin{eqnarray}
R_{i} &=& [1.5^{i-2}, 1.5^{i-1}) R_{\oplus}
\end{eqnarray}
This implies the following bin edges: [0.67, 1.0, 1.5, 2.3, 3.4, 5.1, 7.6, 11, 17,...] $R_{\oplus}$.

The $j^{th}$ bin in the planet period is defined as:
\begin{eqnarray}
P_{j} &=& 10~.~[2^{j-1}, 2^{j}) days
\end{eqnarray}
This implies the following bin edges:
[10, 20, 40, 80, 160, 320, 640,...] days

Data and models from peer-reviewed publications (Petigura et al. 2013, ForemanMackey et al. 2014, Burke et al. 2015, 
Traub 2015, Dressing \& Charbonneau 2013) were integrated over the standard grid, and supplemented by several 
unpublished tables from the 2015 Kepler ``hack week'' which were based on Q1-Q17 DR24 catalog, Kepler completeness 
curves, and data products at the time.

 However, for our current work, we did not use the SAG13 standard grid mentioned above because the SAG13 grid does not 
represent the condensation sequences described in the previous section. Instead, we took the stellar fluxes from Table 1 
where species condensation happens, and calculated the corresponding orbital periods based on the stellar mass 
$(0.965)$ M$_{\odot}$ and luminosity $(0.95)$ L$_{\odot}$ described in footnote 5. 
It should be noted that the SAG-13 grids are available for different spectral types.  Herein, our work focuses on the G dwarf population and employs the corresponding grids.

 The SAG13 submissions were then processed as follows. First, within each spectral type, the sample geometric mean 
$(\mu_{i,j})$ and variance ($\sigma_{i,j}^{2}$) was computed in each ($i,j$)-th bin of the period-radius grid, across the 
different submissions. The mean values $\mu_{i,j}$ formed a ``baseline'' table of occurrence rates. ``Optimistic'' and 
``pessimistic'' tables were also defined by using the $\mu_{i,j} \pm 1 \sigma_{i,j}$ values for each ($i,j$)-th bin.

SAG13 then fit a piecewise power law to the ``pessimistic'', ``baseline'', and ``optimistic'' combined tables. 
The power law had the following form:

\begin{eqnarray}
\frac{\partial^{2}N(R,P)}{\partial lnR~\partial lnP} &=& \Gamma_{i}R^{\alpha_{i}}P^{\beta_{i}}
\label{fit}
\end{eqnarray}

  For optimistic case, $\Gamma_{i} = [1.06, 0.78], \alpha_{i} = [-0.68, -0.82], \beta_{i} = [0.32, 0.67]$; for pessimistic case
$\Gamma_{i} = [0.138, 0.72], \alpha_{i} = [0.277, -1.56], \beta_{i} = [0.204, 0.51]$; for baseline, $\Gamma_{i} = [0.38, 0.73], \alpha_{i} = [-0.19, -1.18], \beta_{i} = [0.26, 0.59]$

 The break between two pieces of the power law was set at 3.4 R$_{\oplus}$ (following Burke et al. 2015), hence the two values
for the coefficients, and a least squares fit 
was performed separately to each of the pieces. Similarly to the mean and variance above, logarithms of occurrence rates were used 
when performing the least squares of log occurrence rates, rather than actual occurrence rates, in order to properly balance the 
effects of small and large occurrence rates. This resulted in ``pessimistic'', ``baseline'', and ``optimistic'' parametric models. 
These models were then integrated across the planet parameter boundaries described in this paper.

It should be stressed that community-sourced data do not represent independent measurements or estimates of scientific quantities, 
so that the SAG13 sample mean and variances should not be interpreted as a formal mean and uncertainty of exoplanet occurrence rates. 
Rather, they simply represent one possible way to measure the state of knowledge as well as the disagreement on the rates within the 
occurrence rate community. In other words, the SAG13 ``pessimistic'', ``baseline'', and ``optimistic'' cases refer to the typical 
pessimistic, average, and optimistic submissions within the SAG13 community survey, rather than formal scientific results. 

Alternative ways of combining SAG13 results are also possible, such as: including only peer-reviewed submissions, including 
submissions based only on different catalogs, removing outliers, etc. A detailed analysis of this is beyond the scope of this paper, 
but as a general rule, combinations tend to fall somewhere between the occurrence rates published in Petigura et al.(2013) and 
Burke et al. (2015) for G-dwarfs, which represent a range of about 4 in the warm rocky planet regime, with a tendency to be closer to 
Burke et al. (2015). For example, the geometric mean combination which we use in this paper is about $25\%$ lower than 
Burke et al. (2015) for warm rocky planets, though it is significantly higher than Petigura et al. (2013). It should also be stressed 
that extrapolation is implied when integrating fitted power laws into cold planets or very small planet sizes, so the numbers in 
those regions remain very unreliable.

 To judge the robustness of the SAG13 occurrence rate estimates, independent occurrence rates were calculated 
using the inverse 
detection efficiency method based on the data from the DR25 catalog.
The occurrence rate per bin, $\eta$, is given by 
\begin{eqnarray}
\eta &=& \frac{1}{n_\star} \Sigma_i^{n_{p}} \frac{1}{\rm comp_{i}}
\label{Eq1}
\end{eqnarray}

Where ${\rm comp}_i$ is the survey completeness evaluated at the radius and orbital period of each planet in the bin, $n_\star$ is the number of stars surveyed, $n_p$ is the number of planets in each bin.
The planet list is taken from \cite{Thompson2017}, using a disposition score cut of 0.9.
The stellar properties are taken from \cite{Mathur2017}, removing giant stars with $logg < 4.2$
 and culling the G dwarfs by selecting stars with $5300$ K $ < T_{eff} < 6000$ K.
The completeness of each star is calculated with \texttt{KeplerPORT} \citep{BurkeCat2017}, and the survey completeness is 
calculated by averaging over all stars that were successfully searched for planets according to the \texttt{timeoutsumry} flag.

The above equation assumes that the vetting completeness (the fraction of planet transit signals 
(TCEs, Threshold Crossing Events) properly classified as planet candidates)
 and reliability 
(the fraction of transiting candidates that are not caused by instrumental artifacts or statistical false alarms)
 are $100 \%$.
The vetting completeness and reliability are very important for small planets, especially at long orbital periods. 
The vetting completeness decreases substantially when one employs a high score cut on the DR25 catalog while the
 reliability approaches $100 \%$.  The net effect is that occurrence rates are likely to be underestimated by 
ignoring both corrections.
 It should be noted that all SAG13 calculations also ignored vetting completeness and reliability. 
We note that more work needs to be done to 
do a reliable occurrence rate calculation.

For regions with no planet detections (low instellation, long period orbits), occurrence rates were estimated with a 
parametric function that is a broken power-law in period and radius. Free parameters were constrained using the 
Exoplanet Population Observation Simulator (\texttt{EPOS}, Mulders et al. in prep). \texttt{EPOS} generates planet 
populations from this parametrized description using a Monte Carlo simulation, and conducts synthetic observations 
using the survey completeness from the DR25 catalog. The synthetic observable populations are compared with the 
observed planet distribution from Kepler in the range $P=[2,400]$ days and $R_{p}=[0.5,8] ~R_\oplus$ , and the 
posterior parameters are estimated using \texttt{emcee} (Foreman-Mackey et al. 2013). Binned occurrence rates are 
calculated by marginalizing the posterior parametric distribution over the bin area, and taking the $50\%$ and $16\%$
 and $84\%$ percentiles for the mean and 1-sigma error, respectively. 

Table \ref{Table3} provides the occurrence rates calculated from Eq.(\ref{Eq1}) for the same bins as in Tables 1 \& 2.
The values are more or less consistent within the uncertainties of SAG13 $\eta_{basl}$ from Table 2. However, the
extrapolations into the cold planet regimes (low instellation fluxes) results in a disagreement between SAG13
values and from Eq. (\ref{Eq1}). This is expected, considering that (1) the cold regimes do not have any planet
detections and any extrapolations are expected to wildly deviate (even between the methodologies) from the
true distribution, and (2) the SAG13 rates are a combination of several individual methodoligies. 

\begin{threeparttable}[h!]
\caption{ Occurrence rates of planets in different boundaries, defined in Table \ref{Table1} classification scheme. The $\eta_{planet}$ values are estimated using SAG-13 occurrence rates. The $\star$ values are based on extrapolation and 
therefore are very uncertain.}
\vspace{0.1 in}
\centering
\begin{tabular}{ |c|c|c|c|c| }
\hline
 Planet Type (Stellar Flux range) & Radius (R$_{\oplus}$) & $\eta_{pess}$ & $\eta_{basl}$ & $\eta_{opt}$\\ 
 \hline
Hot rocky (182-1.0) & 0.5-1.0 & 0.22 & 0.67 & 2.04 \\ 
Warm rocky (1.0-0.28) & 0.5-1.0 & 0.09$^{\star}$ & 0.30$^{\star}$ & 1.04$^{\star}$ \\ 
Cold rocky (0.28-0.0035) & 0.5-1.0 & 0.50$^{\star}$ & 1.92$^{\star}$ & 7.61$^{\star}$\\ 
Hot super-Earths (187-1.12) & 1.0-1.75 & 0.21 & 0.47 & 1.04 \\ 
Warm super-Earths (1.12-0.30) & 1.0-1.75 & 0.087& 0.21 & 0.54 \\ 
Cold super-Earths (0.30-0.0030) & 1.0-1.75 & 0.50$^{\star}$ & 1.42$^{\star}$ & 4.14$^{\star}$\\ 

Hot sub-Neptunes (188-1.15) & 1.75-3.5 & 0.29 & 0.48 & 0.79\\ 
Warm sub-Neptunes (1.15-0.32) & 1.75-3.5 & 0.12 & 0.22 & 0.41 \\ 
Cold sub-Neptunes (0.32-0.0.0030) & 1.75-3.5 & 0.77$^{\star}$ & 1.63$^{\star}$ & 3.52 $^{\star}$\\ 

Hot sub-Jovians (220-1.65) & 3.5-6.0 & 0.05 & 0.07 & 0.12 \\ 
Warm sub-Jovians (1.65-0.45) & 3.5-6.0 & 0.04 & 0.07 & 0.13 \\ 
Cold sub-Jovians (0.45-0.0030) & 3.5-6.0 & 0.58$^{\star}$ & 1.35$^{\star}$ & 3.19$^{\star}$ \\ 

Hot Jovians (220-1.65) & 6.0-14.3 & 0.028 & 0.056 & 0.11 \\ 
Warm Jovians (1.65-0.40) & 6.0-14.3 & 0.023 & 0.053 & 0.12 \\ 
Cold Jovians (0.40-0.0025) & 6.0-14.3 & 0.34$^{\star}$ & 1.01$^{\star}$ & 3.07$^{\star}$ \\ 

 \hline
\end{tabular}
\label{Table2}
\end{threeparttable}

\begin{threeparttable}[h!]
\caption{ Occurrence rates calculated from Eq.(\ref{Eq1}). Comparing with the $\eta_{basl}$ values 
from Table \ref{Table2} from SAG-13 occurrence rates, the SAG13 values are more or less consistent with
the values given below within the uncertainties. We use the $\eta_{basl}$ values from Table \ref{Table2}
to calculate the exoplanet yield estimates in section 2.2. As with Table \ref{Table2}, the $\star$ values
are extrapolated.}
\vspace{0.1 in}
\centering
\begin{tabular}{ |c|c|c| }
\hline
 Planet Type (Stellar Flux range) & Radius (R$_{\oplus}$) & $\eta$ \\ 
 \hline
Hot rocky (182-1.0) & 0.5-1.0 & 0.552$^{+0.195}_{-0.150}$ \\ 
&&
\\
Warm rocky (1.0-0.28) & 0.5-1.0 & 0.215$^{+0.148}_{-0.099}$$^{\star}$ \\ 
&&
\\
Cold rocky (0.28-0.0035) & 0.5-1.0 & 1.09$^{+1.48}_{-0.755}$$^{\star}$ \\ 
&&
\\
Hot super-Earths (187-1.12) & 1.0-1.75 & 0.374$^{+0.068}_{-0.056}$  \\ 
&&
\\
Warm super-Earths (1.12-0.30) & 1.0-1.75 & 0.145$^{+0.071}_{-0.061}$ \\ 
&&
\\
Cold super-Earths (0.30-0.0030) & 1.0-1.75 & 0.78$^{+0.86}_{-0.52}$$^{\star}$ \\ 
&&
\\
Hot sub-Neptunes (188-1.15) & 1.75-3.5 & 0.356$^{+0.049}_{-0.047}$ \\ 
&&
\\
Warm sub-Neptunes (1.15-0.32) & 1.75-3.5 & 0.147$^{+0.058}_{-0.057}$  \\ 
&&
\\
Cold sub-Neptunes (0.32-0.0.0030) & 1.75-3.5 & 0.85$^{+0.88}_{-0.57}$$^{\star}$  \\ 
&&
\\
Hot sub-Jovians (220-1.65) & 3.5-6.0 & 0.113$^{+0.019}_{-0.018}$  \\ 
&&
\\
Warm sub-Jovians (1.65-0.45) & 3.5-6.0 & 0.051$^{+0.021}_{-0.020}$  \\ 
&&
\\
Cold sub-Jovians (0.45-0.0030) & 3.5-6.0 & 0.279$^{+0.31}_{-0.18}$$^{\star}$  \\ 
&&
\\
Hot Jovians (220-1.65) & 6.0-14.3 & 0.004$^{+0.011}_{-0.004}$  \\ 
&&
\\
Warm Jovians (1.65-0.40) & 6.0-14.3 & 0.002$^{+0.004}_{-0.001}$  \\ 
&&
\\
Cold Jovians (0.40-0.0025) & 6.0-14.3 & 0.008$^{+0.031}_{-0.007}$$^{\star}$  \\ 
&&
\\

 \hline
\end{tabular}
\label{Table3}
\end{threeparttable}

\begin{figure}
\includegraphics[width=0.85\textwidth]{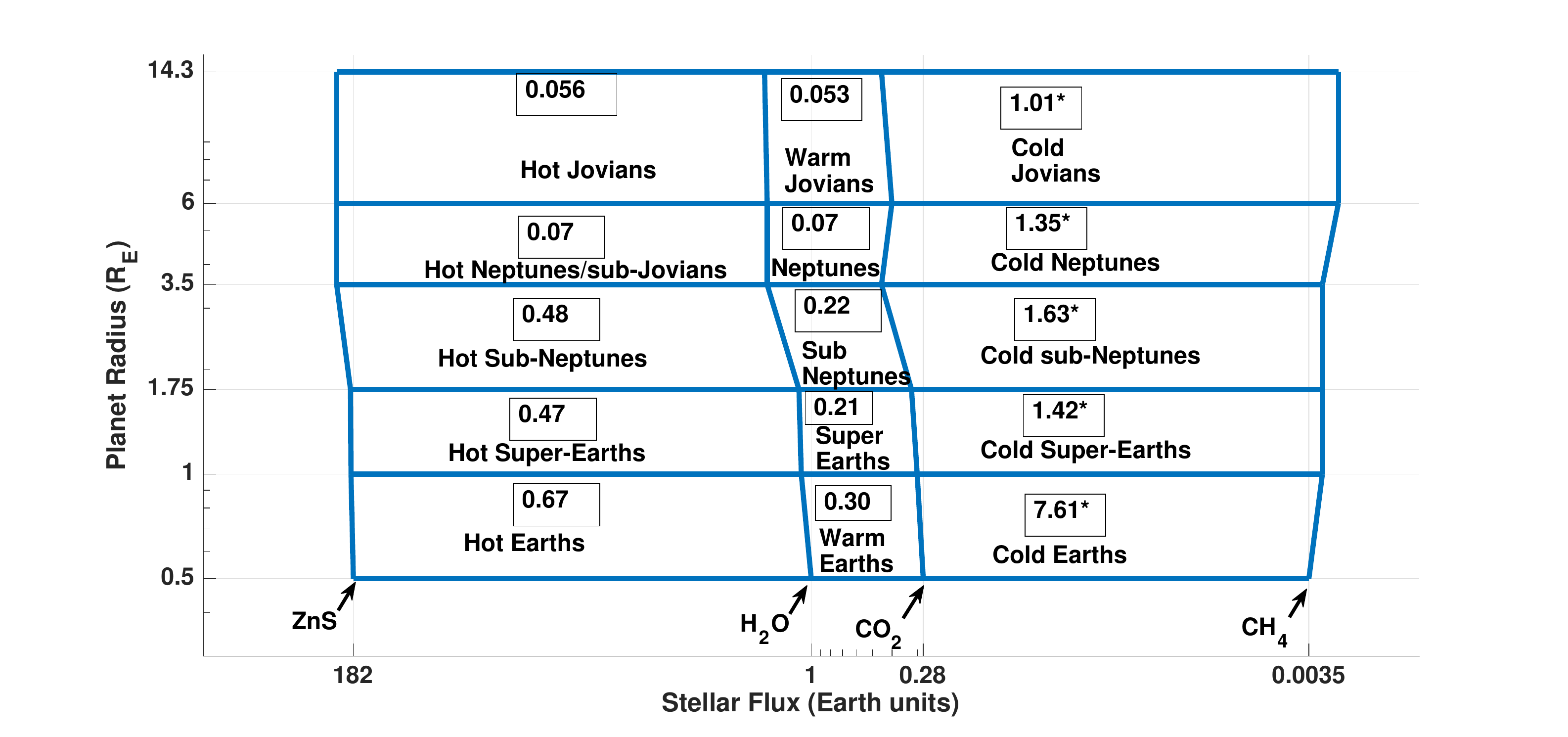}
\caption{ Planet occurrence rate estimates from SAG13 baseline analysis (see Table \ref{Table2}) 
as a function of incident flux and planetary radius, assuming a star with
 $L = 0.95 L_{\odot}$ and $M = 0.965M_{\odot}$. The boundaries of the 
 boxes represent the regions where different chemical species are condensing in the atmosphere of that particular size 
planet at that stellar flux, according to equilibrium chemistry calculations. The radius division is from 
\cite{Fulton2017} for super-Earths and sub-Neptunes, and from \cite{ChenKipping2017} for the upper limit on Jovians.
 The `$\star$' values are based on 
extrapolation and therefore are very uncertain. See $\S 2.1$ for more details.}
\label{occr}
\end{figure}

Integrating the SAG13 parametric model across the bin boundaries defined in Table \ref{Table1} gives the occurrence rates in 
Fig \ref{occr}, where we have assumed $L = 0.95 L_{\odot}$ and $M = 0.965 M_{\odot}$. Each class of planet has an 
occurrence that is a mixture of astrophysical effect and an observational bias. For example, even though it is easier 
to detect giant planets in close orbits, their occurrence rate is comparatively smaller ($0.05$) than close-in 
sub-Neptune or terrestrial-size planets ($\sim 0.48$, respectively). The implication is that hot, giant planets are 
likely less in number.  

 The trend in Fig. \ref{occr} indicates that the occurrence rate generally increases, from larger planets to the smaller ones in any particular bin. We can define the inner edge of the habitable zone (HZ) as the boundary where $\h2o$ starts 
condensing in a terrestrial planet’s atmosphere, and the outer edge of the habitable zone as $\co2$ condensation 
boundary \citep{Abe2011}. Within this zone, it appears that the terrestrial size planets have 
higher occurrence rates ($0.2 - 0.3$) than compared to either Jovians (0.053) or Neptunes (0.07) planets. 
However, it should be noted that the 
occurrence rate of terrestrial planets in this regime is severely restricted by low number statistics. 

\subsection{Mission yield estimates}
With the planet categorization scheme and associated occurrence rates described above, we estimated the exoplanet yields for each type of planet using the yield optimization code of \cite{Stark2015}. Briefly, this code works by simulating the detection of extrasolar planets around nearby stars over the lifetime of a mission.  To do so, it distributes a large number of synthetic planets around each nearby star, sampling all possible orbits and phases consistent with the planet definition, illuminates them with starlight, calculates an exposure time for each planet given a set of assumptions about the instrument and telescope, and determines the fraction of planets that are detectable within a given exposure time (i.e., the ``completeness"). The code optimizes the exposure time of each observation, as well as which stars are observed, the number of observations to each star, and the delay time between observations, to maximize the yield for a given type of planet.

We first ran the yield code to define the set of observations that maximized the yield of exoEarth candidates. 
 We adopted the same baseline mission parameters defined in Table 3 of Stark et al. (2015) with exception to the OWA, 
which increased from 15 $\lambda$/D to 30 $\lambda$/D, the contrast for spectral characterization which improved from 
$5\times 10^{-10}$ to $1\times 10^{-10}$, the spectral resolution which increased from R=50 to R=70, and the SNR required 
for spectral characterization which increased from 5 to 10 per spectral channel.
 We also adopted a new definition for exoEarth candidates.  We distributed exoEarth candidates across the 
\cite{Kopp2014} conservative HZ, which ranges from 0.95 - 1.67 AU for a solar twin. The semi-major axis distribution 
followed the analytic SAG13 occurrence rate fits.  ExoEarth candidates ranged from 0.5 — 1.4 Earth radii, with the lower 
radius limit set by $0.8*(a * (L_{\star})^{1/2})^-0.5$, a reasonable limit following the work of \cite{ZC2017}.
  All exoEarth candidates were assigned a flat geometric albedo of 0.2. In this paper, the exoEarths are used solely to 
optimize an observation plan. We do not report on the yield of these exoEarth candidates, focusing instead on the yields 
of the classes of planets defined in section 2 when following such an observation plan.
 We then locked this set of observations in place and re-ran the yield code to 
calculate the yield of each planet type discussed above, simply by changing the planet’s input parameters each time. 
For each planet type, we distributed planet radii and orbital period according to the SAG13 distribution. We assumed 
Lambertian phase functions for all planets.

To calculate the brightness of a given planet, we must also know the planet’s albedo. The actual distribution of exoplanet 
albedos is unknown. So for this study, we simply assigned each planet type a single reasonable albedo. We adopted a 
wavelength-independent geometric albedo of 0.2 for rocky planets and 0.5 for all other planets. 

\begin{figure}
\subfigure[]{
\label{fourm}
\includegraphics[width=.45\textwidth]{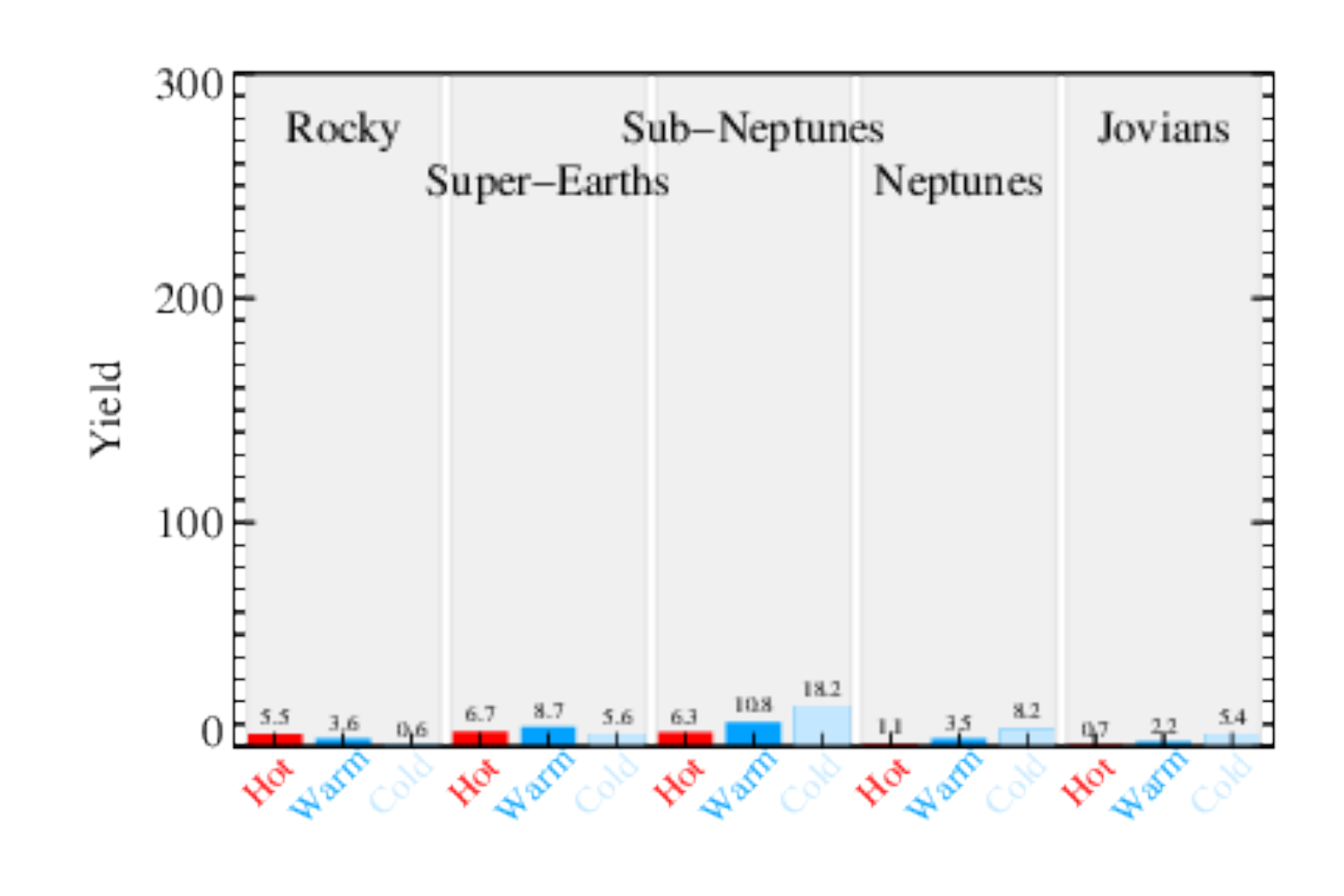}
}
\subfigure[]{
\label{eightm}
\includegraphics[width=.45\textwidth]{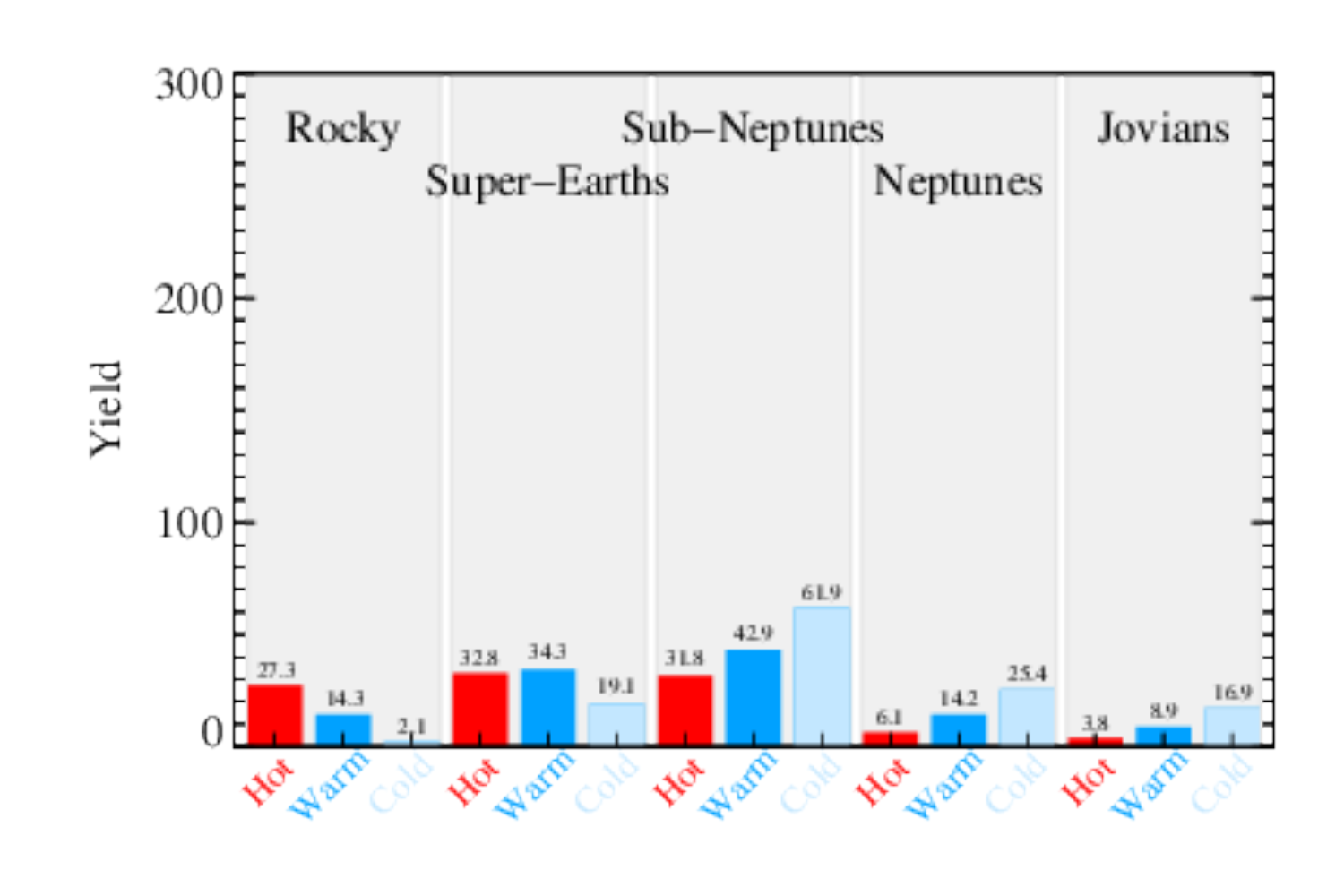}
}
\begin{center}
\subfigure[]{
\label{sixteenm}
\includegraphics[width=.45\textwidth]{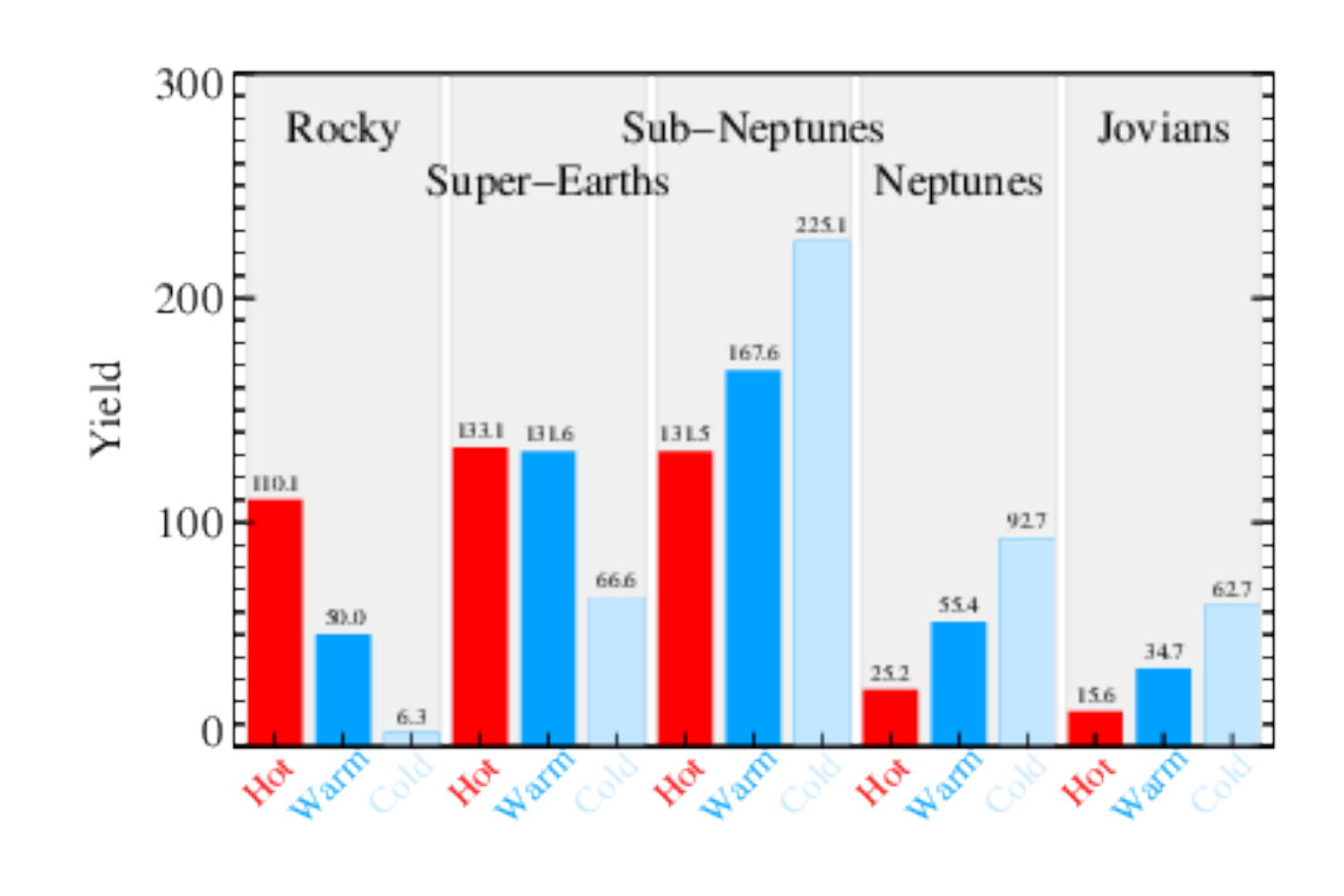}
}
\end{center}

\caption{ Expected number of exoplanets observed (y-axis) for the baseline occurrence rates in each planet category 
(rocky, super-Earths, sub-Neptunes, sub-Jovians and Jovians) for hot (red), warm (blue) and cold (ice-blue) incident 
stellar fluxes shown in Table 1 and Fig.2. The telescope sizes are (a) 4m, (b) 8m and (c) 16m. The occurrence rates,
as well as yield estimates, ignore multiplicity and the planet categories were all treated as effectively independent.}
\label{hist}
\end{figure}

To calculate an expected yield, we must also assign each planet type an occurrence rate. Table \ref{Table2} lists the occurrence rates obtained from \S\ref{Occurrence} in each bin of planet size and planet type (hot/warm/cold). The histogram plot in Fig.\ref{hist} visualizes the total scientific impact of the habitable planet candidate survey. The y-axis gives the expected total numbers of exoplanets observed (yields), which are also given by the numbers above the bars. By ``expected'', we mean the most probable yield after many trials of an identically executed survey. Three sizes of exoplanets are shown, consistent with Table 1. For each planet size, three incident stellar flux classes are shown: hot (red), warm (dark
blue), and cold (ice blue). The boundaries between the classes correspond to the temperatures where metals, water vapor, and carbon dioxide condense in a planet's atmosphere. The warm bin is not the same as the habitable planet candidate bin,
 as it is likely too generous.

\begin{figure}
\subfigure[]{
\label{eight_y}
\includegraphics[width=.85\textwidth]{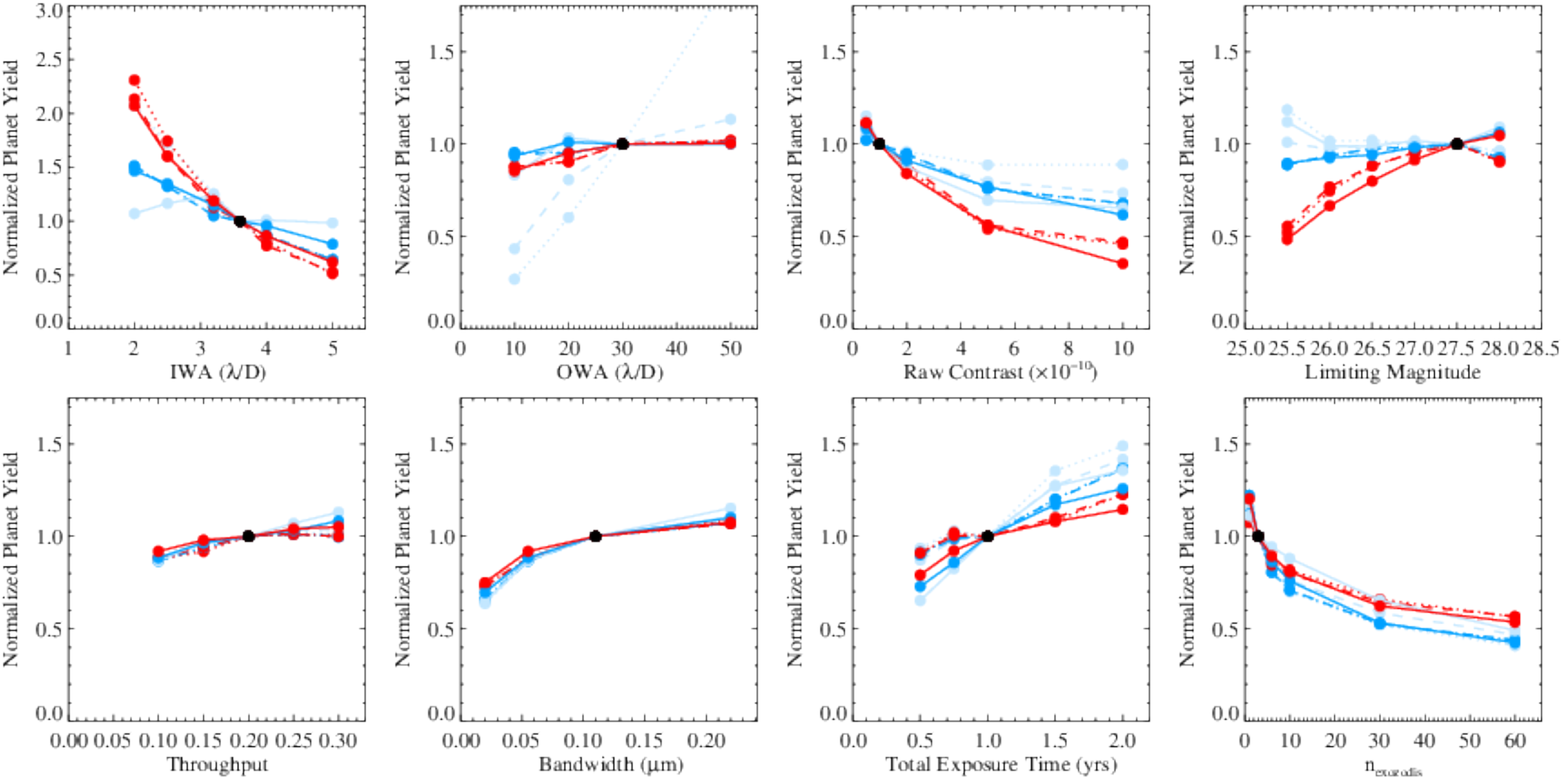}
}
\subfigure[]{
\label{twelve_y}
\includegraphics[width=.85\textwidth]{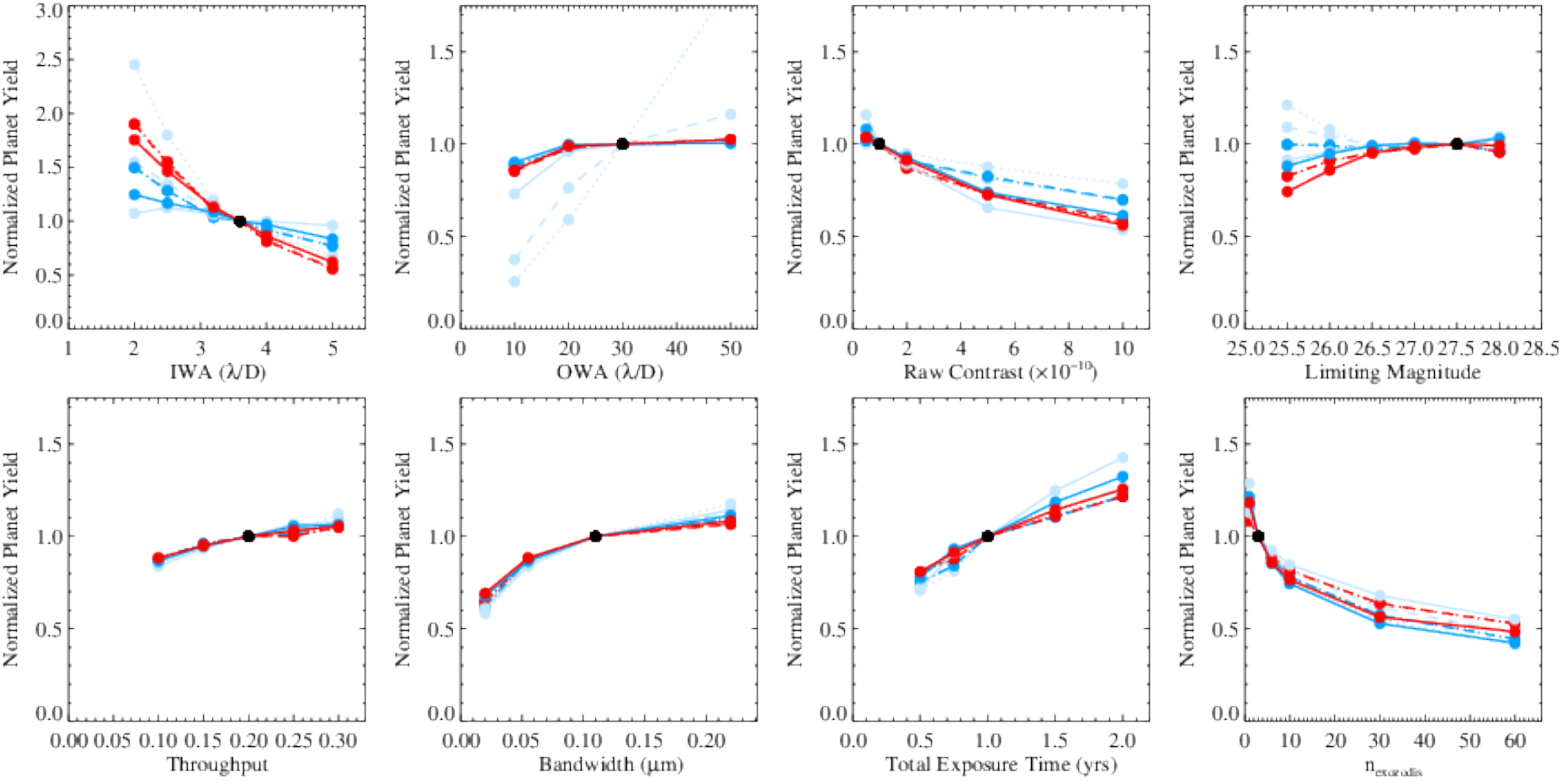}
}
\caption{ Yield for each planet type when deviating from the baseline mission by varying one parameter at a time. 
The top panel is for a 4m mirror size and bottom is for 8 meter. Solid, dashed, and dotted lines correspond to rocky, 
sub-Neptune, and Jovian planet types. Color scheme is the same as Figure 3.}
\label{yield_sensitivity}
\end{figure}
\begin{figure}
\includegraphics[width=.85\textwidth]{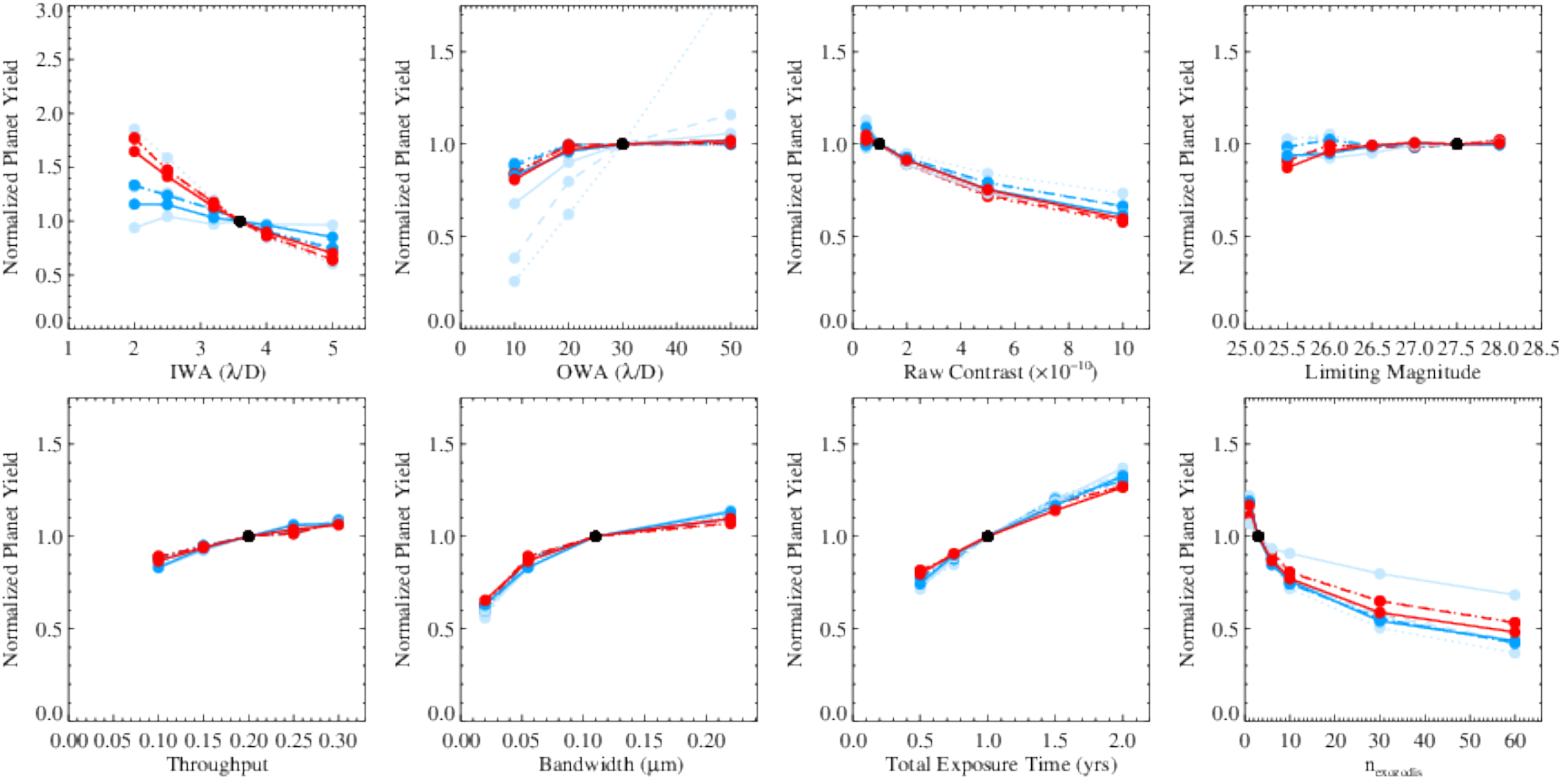}
\caption{Same as Fig. \ref{yield_sensitivity}, but for a 16m mirror size.} 

\label{sixteen_y}
\end{figure}

We also calculated each planet's yield when deviating from the baseline mission parameters. Figs. \ref{yield_sensitivity} \& \ref{sixteen_y} 
show the sensitivity of each planet's yield to changes in a single mission parameter.  Each yield curve has been 
normalized to unity at the value of the baseline mission.
As expected, the yield of hot planets is more sensitive to the IWA than cold planets, and the yield of cold planets is 
more sensitive to OWA than hot planets.  However, surprisingly, the yield of cold Jupiters is quite sensitive to IWA, 
suggesting that an observation plan optimized for the detection of exoEarths will typically detect cold Jupiters in 
gibbous phase near the IWA.  
We note that in general, larger apertures are less sensitive to changes in 
mission parameters than smaller apertures.

\section{Discussion}

In this section, we discuss caveats to our classification scheme, and its relevance to future missions that plan to detect and/or characterize extrasolar planets.

\subsection{Caveats to the classification scheme}
 The boundaries discussed in earlier sections are made out of the necessity for creating a single classification scheme that applies to all planets, and that can translate current planet obervations into predictions of future planet yields. No single scheme will be able to properly capture the complex interactions between myriad planetary processes in exoplanet atmospheres. An analogy with the habitable zone here is useful: The habitable zone is not a good means of determining the habitability of a single planet. Instead, it is best used to understand how many potentially habitable worlds a given mission may be able to probe for evidence of habitability. Similarly, the classification scheme discussed in this paper is probably not the best means for determining whether a single world has a certain combination of clouds decks or upper atmospheric composition. Ultimately, that should be determined by specific observations of such worlds. What this scheme \textit{is} useful for is to help understand the diversity, and the number, of worlds that can be expected from future missions and ground-based observatories. 

Even given the above overall caveat, more specific caveats exist. Much of the work upon which this classification scheme is based is relatively new, and represents an area of rapidly evolving thought. This means that the resulting classification scheme will also have to evolve as these theories are better elucidated, tested, and refined. Ultimately, when future exoplanet characterization missions and ground-based observatories become operational, they will provide the tests of the various hypotheses contained in the drawing of the boundaries shown in Fig. 2. This again draws similarities to the habitable zone, which has evolved over the years, and will ultimately be determined by missions. But until such observations are made, this scheme (and the HZ) will enable predictions of how many such worlds we will be able to eventually epxlore in detail.

An example of this rapid evolution is the work by myriad groups on the mass-radius relationship of planets \citep{WM2014,Rogers2015, Wolfgang2016, ChenKipping2017} Clearly, this area of research is rapidly evolving - and further data on 
this relationship is anticipated with continues ground-based observations of planets around M dwarfs combines with TESS 
observations of the same target. Thus, we anticipate changes to the boundaries selected in this study as these new data and models are incorporated into this line of research. We also anticipate that these data will allow for the study of how these dividing lines are a function of instellation received by the planet. If that is the case, the "horizontal" lines in our classification scheme (Figure 2) would instead be diagonal. We used horizontal lines here as a first-order determination of this scheme and because deriving the slope of those lines empirically or theoretically is well beyond the scope of this work. This is something that may be true more generally - the lines drawn in our scheme may all be diagonal, as we think instellation could impact planet size categories and that planet size could impact the instellations at which various cloud decks form.

A second category of caveat is the lack of consideration of many planetary processes and planet/star properties in this study. Kinetic chemistry, atmospheric circulation, internal heat generation, different star types, and differing bulk planet composition are not explicitly considered in this study. Each of these processes could affect the boundaries considered here.

Our boundaries have been determined at equilibrium. However, the distribution of chemical constituents in planetary 
atmospheres can be strongly affected by the so-called chemical quenching, in which as the material moves, the temperature 
and pressure within the gas mixture change and chemical reactions may become slower, potentially reaching a certain state 
for which the chemical abundances are `frozen' \citep{Madhu2016}. Also, photochemistry could also impact the ability of a 
condensate to form.
This is most evident in the potential for secondary aerosols that form when photochemical byproducts cause supersaturation and condensation in an atmosphere. This process occurs on many solar system worlds \citep{Courtin2005, Waite2007, Gladstone2016},
 is thought to have occurred on Archean Earth \citep{Arney2016}, and likely occurs on exoplanets \citep{Moses2011, Moses2014}.
 However, these processes themselves  will likely represent a significant overprint over the transitions that are proposed here, where within one of the planet categories we propose, planets that are closer to their host star will have a gradual and increasingly important contribution from non-equilibrium chemistry. 
But others may also exist, including the potential for other photochemical byproducts to form in an atmosphere and be evidenced in the planetary spectrum. However, we expect these will be secondary in importance compared to the optical depth and cold-trapping effects of discrete cloud layers.

The specific properties of the planet and star are also not considered here. Atmospheric circulation, internal heat generation, and the stellar energy distriburion (SED) are also ignored here. These could all impact the specific position of the boundaries we propose. Changing the SED can lead to changes in the planetary albedo, which would affect the position of these boundaries just as they affect the habitable zone \citep{Segura2003, Shields2013, Shields2014, Yang2013, Yang2014a, Kopp2014, Kopp2016, Wolf2017}. Greater internal heat generation, or decreased convection could lead to a suppression of cloud formation. These are not independent variables, as increased heat generation should lead to increased convection and the SED could also impact circulation as it impacts the altitude-dependent deposition of energy in the atmosphere. Although these properties and processes are not included here, they could be in future papers, just as the consideration of the habitable zone now includes consideration of internal heat sources \citep{Barnes2013, HM2015}, atmospheric circulation and convection. Finally, we do not consider the impact of the bulk chemical composition of the planet. Clearly, a planet that is extremely carbon poor will be less likely to condense  CO$_{2}$ or CH$_{4}$.

Lastly, we note that the simulations of the scheme in this paper considers a limited parameter space, due to its focus on a near-term problem faced by the exoplanet community. Given the ongoing studies of  flagship missions that propose to directly image and spectroscopically characterize extrasolar planets - HabEx and LUVOIR - we require an ability to discuss and compare the exoplanet yields from such missions. This will also help us understand how these missions will complement the discoveries that will be made by already-planned ground-based and space-based observatories. The focused utility shown in this manuscript comes at the expense of other applications of the proposed classification scheme. It could also be used to simulate the yields from future transit spectroscopy missions such as JWST or CHEOPS. It also helps us understand how detection missions such as TESS, PLATO, and  WFIRST will provide complementary discoveries to past exoplanet detections, and increase the total expected diversity of known planets. These are all worthy applications of the scheme we propose, but we save these applications for future manuscripts.

\subsection{Application of classification scheme to future space-based direct imaging missions}

These estimates of the abundances of different planet types allow for projections of the yields of future exoplanet missions. As mentioned in the introduction, two such missions are under study in advance of the next astrophysics Decadal Survey:  HabEx and LUVOIR. Most of the discussion of these missions - and past direct imaging studies - has focused on their ability to find and characterize rocky planets in the habitable zones of nearby stars. However, the observations enabled by such missions would also bring an ability to observe other kinds of worlds. All planets between the inner working angle and outer working angle of the starlight suppression at the time of observation can be observed. And many of these worlds will be brighter than rocky planets in the habitable zone. Therefore, even observation strategies that are optimized for maximizing the yield of rocky planets in the habitable zone will also yield observations of a considerable diversity of worlds. 

None of these yield estimates should be taken as simulations of the yields of the architectures the HabEx and LUVOIR teams are studying. For one, these simulations hold everything except telescope diameter constant, as a way to demonstrate how planet diversity scales with telescope size and as a way to show end-member yields across the range of mission sizes being explored (For a more extensive options of varying quantities other than the telescope diameter, see here: \url{http://jt-astro.science:5100/multiplanet_vis}). More importantly, the two studies will explore four different architectures which may have different instrument properties, assume different levels of technological development, and different starlight suppression techniques. None of those nuances are considered here. We refer the reader to the eventual reports from these teams for better estimates of the yields from these missions. What we present here should instead be considered an example of how these planet categories can be useful to such yield simulations.

Our simulations predict that a 4m-class mission 
would observe a great diversity of worlds (Figure \ref{fourm}). At that scale, a mission whose observations 
are designed to maximize the yield of potential Earths will also yield the detection and characterization of all of the 
planet types discussed here, with the exception of hot Jupiters. Hot Jupiters are not observed by a 4m-class mission 
because the tight inner working angle of the 4m mission, and because of the low abundances of hot Jupiters (see Table 2)
. But a total of up to $90$ planets are observed, including up to $12$ rocky/super-Earths planets in the habitable zone. 
The diversity of these simulations will allow tests - albeit on a small sample size within each bin - of the planet 
bins proposed here.

An 8m-class mission yields even more planets (Figure \ref{eightm}). We predict that class of mission would observe over 
300 planets, including at least a few planets in each of the planet classes proposed in this manuscript.  Although 
neither LUVOIR nor HabEx is explicitly considering an 8m mission, this size sits at the dividing line between the two 
studies and represents an approximation of what a large HabEx or a small LUVOIR could enable. 

The larger versions of LUVOIR (Fig. \ref{sixteenm}) bring the ability to not only observe planets, but to test the occurrence of 
different features 
within each of the planet bins. It would observe dozens of each planet type, providing larger sample sizes which enables to study each 
planet type as a population. As shown by Stark et al. 2014, a sample of 30 planets of a given type would allow us to be sensitive to any
 feature that has at least a 10\% chance of occurring. For this class of mission, there would be a total of 
up to $1000$ worlds. This includes several hundred
Neptune-size planets with orbits that likely cause water clouds - but not carbon dioxide clouds - to form. The least populated bins are 
for hot Jupiters, which only contains $\sim 15$ detections, and for ``cold rocky'' which would have $\co2$ clouds, 
which contains only few 
detections. But all other bins have at least 50 detections. As mentioned above, 30 planets of a given type would allow 
us to be sensitive 
to any feature that has a $\sim 10\%$ chance of occurrence within that planet type (Stark et al. 2014).
 It is this large number of observation that would allow the myriad hypotheses contained in this manuscript to be 
tested. For example, the presence/absence of various cloud types could be plotted as a function of the energy incident 
at the top of exoplanet atmospheres. And the absence of absorption features associated with cold-trapping could be 
measured in each of these bins. This logic has been applied to habitable planets before \citep{Stark2014}; here we 
demonstrate that it also applies to planets beyond the habitable zone.

\section{Conclusions}
\label{conclusions}

NASA concept mission studies that are currently underway,  like LUVOIR and HabEX, are expected to discover a large 
diversity of exoplanets. These larger missions would provide large enough sample sizes that we could study each planet 
type in the context of its ``relatives'' in the Solar system or currently known exoplanets.
We present here a classification scheme for exoplanets in planetary radius and stellar flux bins, based on chemical species' condensation sequences in planetary atmospheres. This chemical behavior of gases is dependent on pressure, temperature and metllicity, and we primarily focus on condensation curves of sphalerite ZnS, $\h2o$, $\co2$ and $\ch4$. The order of condensation of these species represent the order in which the boundaries of our classification scheme are defined. We then calculated the occurrence rates of different classes of exoplanets within these boundaries, and estimated the direct imaging mission yields for various telescope sizes. While the main focus of future flagship exoplanet direct imaging missions is to characterize a habitable world for bio-signatures, the missions will also have the ability to observe other kinds of planets within the system. Therefore, distinguishing features that separate planet categories based on current observables (planet radius and incident stellar flux), and a scheme to illustrate these categories, is essential in calculating the expected direct imaging mission yields, and correspondingly choosing an optimal observational strategy.

        \acknowledgements

The authors would like to thank an anonymous reviewer for their constructive comments. We also thank Aki Roberge,
Mark Marley and Eric Lopez for providing comments on the work and the manuscript. The authors would like to 
gratefully acknowldege the following colleagues who have contributed to SAG-13 community occurrence rate project: 
Chris Burke, Joe Catanzarite, Courtney Dressing, Will Farr, B. J. Fulton, Daniel Foreman-Mackey, Danley Hsu, 
Erik Petigura, Wes Traub.

We would also like to thank the NASA's {\it Kepler} 
mission for organizing the ``hack week'' from October 13 - 15 2015, that provided a forum for project scientists and community 
researchers to interact, and led to significant input into this manuscript.
     R. K and S. D. G. gratefully acknowledge funding from NASA Astrobiology Institute's  Virtual
       Planetary Laboratory lead team, supported by NASA under cooperative agreement
       NNH05ZDA001C. E.H. was supported by an appointment to the NASA Postdoctoral Program at NASA Goddard Space Flight Center, administered by Universities Space Research Association through a contract with NASA.

\end{document}